\documentclass[aps,pre,twocolumn,amsmath,amssymb,amsfonts,floatfix,superscriptaddress,showkeys,showpacs]{revtex4-1}
\usepackage{dcolumn}  
\usepackage{multirow}
\usepackage[T1]{fontenc}
\usepackage[]{inputenc} 
\usepackage{dsfont}
\usepackage{verbatim}
\usepackage{bm} 
\usepackage{hyperref} 
\usepackage{cleveref} 
\usepackage{floatrow}
\usepackage{natbib}
\usepackage{graphicx}
\usepackage{epstopdf}
\usepackage{ifpdf} 
\usepackage{epsfig}
\usepackage{color}
\usepackage[usenames,dvipsnames]{xcolor}
\usepackage[caption=false]{subfig}
\usepackage[export]{adjustbox}
\usepackage[colorinlistoftodos]{todonotes}

\newcommand{\rmd}{{\mathrm{d}}}
\newcommand{\rme}{{\mathrm{e}}}
\newcommand{\icomplex}{\dot\iota}
\newcommand{\kBT}{k_{\mathrm{B}}T}
\newcommand{\kB}{k_{\mathrm{B}}}
\newcommand{\tr}{{\mathrm{tr}}}
\usepackage{cancel}

\begin{document}

\title{Surface alignment disorder and pseudo-Casimir forces in smectic-A liquid crystalline films}

\author{Fahimeh \surname{Karimi Pour Haddadan}}\thanks{f.karimi@khu.ac.ir}
\affiliation{Faculty of Physics, Kharazmi University, Tehran 15815-3587, Iran}
	
\author{Ali \surname{Naji}}
\affiliation{School of Physics, Institute for Research in Fundamental Sciences (IPM), Tehran 19395-5531, Iran}

\author{Rudolf \surname{Podgornik}}
\affiliation{School of Physical Sciences and Kavli Institute for Theoretical Sciences, University of Chinese Academy of Sciences, Beijing 100049, China}
\affiliation{CAS Key Laboratory of Soft Matter Physics, Institute of Physics, Chinese Academy of Sciences, Beijing 100190, China}

\begin{abstract}
Random (disordered) components in the surface anchoring of the smectic-A liquid crystalline film in general modify  the thermal pseudo-Casimir interaction. Anchoring disorder of the quenched type is in general decoupled from the thermal pseudo-Casimir force and gives rise to an additional disorder-generated interaction, in distinction to the annealed disorder, whose effect on the pseudo-Casimir force is non-additive. We consider the effects of the surface anchoring disorder by assuming that one of the substrates of the film is contaminated by a disorder source, resulting in a Gaussian-weighted  distribution of the preferred molecular  anchoring orientation (easy axes) on that substrate, having a finite mean and variance or, more generally, a homogeneous in-plane, two-point correlation function. We show that the presence of disorder, either of the quenched or annealed type, leads to a significant  reduction in the magnitude of the net thermal fluctuation force between the confining substrates of the film. In the quenched case this is a direct consequence of an additive free energy dependent on the variance of the disorder, while in the annealed case, the suppression of the interaction force can be understood based on a disorder-renormalized, effective anchoring strength.  
\end{abstract}


\maketitle

\section{Introduction}

Casimir interactions, in general denoting quantum as well as thermal fluctuation induced interactions of the electromagnetic fields confined by hard or soft substrates \cite{Woods}, play a fundamental role in materials science, atomic and molecular physics, condensed matter physics, high energy physics, as well as chemistry and biology \cite{Parsegian,Bordag,French}.  Despite their 
venerable  nearly 70-year history \cite{Casimir}, there is no shortage of new, fundamental insights be it from the theoretical \cite{Rahi,Woods} as well as the experimental sides 
\cite{Somers,zumer}. One of the fundamental insights was the realization by Fisher and de Gennes  \cite{Fisher}, later analyzed in detail by Dietrich and coworkers \cite{Dietrich},  that a pseudo-Casimir effect can be expected in any confined field theory close to a critical state, e.g.,  critical density fluctuations in a liquid state theory, as well as director fluctuations or layer spacing fluctuations in a liquid crystal theory \cite{Ajdari, Kardar}. The liquid-crystalline order \cite{de-gennes} is usually established at non-zero temperatures $T\neq 0$ and thus the thermal energy $\kB T$ (with $\kB$ the Boltzmann constant) can easily drive fluctuations in the relevant order parameters. As a result, these thermal fluctuations of liquid crystalline order in confined geometries  may induce a macroscopic pseudo-Casimir interaction analogue, in some respects similar but in others differing substantially from the original electromagnetic Casimir interactions \cite{Woods}. 

However, the universal thermal disorder is not the only possible source of fluctuations and fluctuation-driven interactions.  Inherent structural disorder, as in the case of surface charge disorder on dielectric interfaces \cite{sndhp}, has been recognized as possibly not only modifying the thermal Casimir interactions in its annealed variety, but even as a source of a new disorder-driven interaction when the disorder is quenched \cite{ddnp}.  Structural disorder can appear also in the case of liquid crystalline systems. The periodic arrays of the nanometer sized layers of a smectic-A liquid crystal confined between two parallel interfaces normal to the layers can easily be subjected to a quenched disordered-surface anchoring and the role of the disorder in this system, being experimentally easily accessible, has already been analyzed in detail \cite{radzihovsky}. The effects of disorder on the pseudo-Casimir interactions in the case of confined nematic liquid crystals have been studied for a nematic liquid-crystalline film bounded by two planar surfaces perpendicular to the nematic axis, one of which imposes a random (disordered) distribution, quenched or annealed, of either the preferred anchoring axis  \cite{faar} or exhibiting a disordered distribution of anchoring energies \cite{fanr}. In the former case, the quenched disorder effects appear additively in the total interaction while for the annealed disorder its effects are non-additive and can be rationalized in terms of  renormalization of the effective surface anchoring parameter, leading to quantitative and qualitative changes, including a change of sign, in the pseudo-Casimir interaction strength. 

Another interesting aside of the thermal pseudo-Casimir effect in the case of liquid crystalline systems and indeed in many other soft matter systems, is that formally it can be reduced to functional integrals of higher derivative field Lagrangians \cite{Dean-FI}. In the case of liquid crystals this turns out to be a second derivative field Lagrangian, but higher orders are in principle also possible. There are several formal aspects of the theory of higher derivative functional integrals that deserve to be studied in more detail.

The {\em transverse pseudo-Casimir force} arises as a result of thermal fluctuations of the liquid crystalline layers of a smectic-A film confined between two planar substrates in a bookshelf geometry, in which the equidistant smectic layers are placed perpendicular to the bounding surfaces \cite{FAR}. The fluctuation-induced interactions within such a cell in the absence of the  disorder were shown to be attractive (for similar, or symmetric, boundary conditions on the two substrates) and depend on the penetration length as well as layer spacing \cite{FAR}. Below we generalize this setup to include quenched and annealed disorder in the easy axis of the surface anchoring and study the disorder-induced modification of the fluctuation interaction force. Effects of quenched disorder, which enters the problem through its variance appear to be additive and act to weaken the attraction between like boundaries as the variance increases.  
Our formalism including the effects of quenched disorder is based on a fixed equilibrium configuration of the order parameter and does not take into account a possible destabilization of this uniform equilibrium configuration of a sm-A upon increase in the disorder. 

We introduce our model in Section~\ref{sec:model}. The formalism to evaluate the free energy of the system through the partition function follows in Section \ref{sec:formalism}. The analytical and numerical results are presented in Section~\ref{sec:force} and 
the concluding remarks in Section \ref{sec:con} summarize  the conceptual background and the salient features of the results.

\section{Model}
\label{sec:model}

A film of smectic-A liquid crystal (sm-A)~\cite{de-gennes} is assumed to be confined between two parallel plates (or substrates) placed at locations $y=0$ and $d$ of a Cartesian coordinate system ${\mathbf r}=({\mathbf r}_{\!_\perp}, y)$, with lateral coordinates ${\mathbf r}_{\!_\perp}=(x,z)$. The sm-A layers are oriented perpendicular to the substrates, with parallel layers of separation  $a_0$, stacked up in the $z$ direction, resembling a bookshelf configuration for the constituent molecules (see Fig. 1 in Ref. \cite{FAR}). Non-similar anchoring conditions are imposed on the two substrates; while an (infinitely) strong anchoring condition is retained on the substrate positioned at $y=0$, a finite-strength anchoring is allowed  on the substrate positioned at $y=d$, with a random (disordered) component, which we shall specify below. 

The (free) energy of the layer displacements $u({\mathbf r}_{\!_\perp}, y)$ in the bulk of the material has the assumed harmonic form
\begin{equation}
{\cal H}_{\mathrm{b}}=\frac{K}{2}\int_{V} \rmd{\mathbf r} \left[\left(\partial_{x}^2 u+\partial_{y}^2 u\right)^2+{B\over K}(\partial_{z}u)^2\right],
\label{bulk}
\end{equation}
where $K$ is the Frank elastic constant corresponding to bending deformations of the layers, $B$ is the compression (dilatation) elastic constant, and $V$ is the volume of the smectic slab~\cite{de-gennes}.  The strong surface anchoring at $y=0$ is implicitly assumed to keep the mean molecular orientation, or the director field  ${\mathbf n}({\mathbf r}_{\!_\perp}, y)$, strictly in the $z$-direction. On the other hand, the finite surface anchoring at $y=d$ stipulates that  ${\mathbf n}$ orients in the direction of the easy axes ${\mathbf e}({\mathbf r}_{\!_\perp})$  on that substrate. This amounts to the surface interaction energy of the Rapini-Papoular form \cite{RP}  as
\begin{equation}
{\cal H}_{\mathrm{s}}=-\frac{W}{2}\int_{\partial V} \rmd{\mathbf r}_{\!_\perp}\left({\mathbf n}\cdot {\mathbf e}\right)^2,
\label{H_s}
\end{equation}
where $W$ is the anchoring strength and the integral runs over the surface area of the bounding substrate at $y=d$
~\cite{RP,radzihovsky1}. 
 
We assume that the mean orientation of the easy axis is along the $z$-direction, i.e., $\langle {\mathbf e}({\mathbf r}_{\!_\perp})\rangle=\hat {\mathbf z}$. We use the gauge $(\partial_{x},\partial_{y})^T u+\delta {\mathbf n}=0$ where $\delta {\mathbf n}=(\delta n_{x}, \delta n_y)$ is the small deviation from the mean orientation $\langle {\mathbf n}\rangle =\hat{\mathbf z}$, to rewrite Eq.~\eqref{H_s} as 
 \begin{equation}
{\cal H}_{\mathrm{s}}=\frac{W}{2}\int_{\partial V} \rmd{\mathbf r}_{\!_\perp}\left[(\partial_{x}u)^2+(\partial_{y}u)^2+2 (e_x \partial_x u+e_y \partial_y u) \right],
\label{third}
\end{equation}
up to the linear order in the disorder field ${\mathbf e}\simeq (e_{x},e_y,1)$. This makes the director to remain locally perpendicular to the layers~\cite{de-gennes,radzihovsky2,lubensky}. 

We assume that the disorder fields  $e_i(x,z)=\{e_x(x,z), e_y(x,z)\}$  are statistically independent and have Gaussian distributions around their zero mean values, 
\begin{equation}
{\cal P}[e_i]={\cal C}\exp^{-{1\over 2}\int \rmd{\mathbf r}_{\!_\perp}\!\rmd{\mathbf r^{\prime}}_{\!_\perp}e_i({\mathbf r}_{\!_\perp} )c^{-1}({\mathbf r}_{\!_\perp}-\mathbf r^{\prime}_{\!_\perp})e_i(\mathbf r^{\prime}_{\!_\perp})}, 
\end{equation}
where ${\cal C}$ is a renormalization constant and $c({\mathbf r}_{\!_\perp}-\mathbf r^{\prime}_{\!_\perp})$ is the in-plane, two-point correlation function of the disorder fields, which we have assumed to have a translationally invariant form in the in-plane directions. 

The thermodynamic free energy of the system follows from averaging the sample free energy over all realizations of the disorder fields as 
${\cal F}_Q
=-\kB T \langle \ln {\cal Z} \rangle$ in the {\em quenched} case, and as ${\cal F}_A=-\kB T\ln \langle {\cal Z} \rangle$ in the  {\em annealed} case, where  $$\langle (\cdots) \rangle = \prod_i\int{\cal D}e_i\, (\cdots){\cal P}[e_i]$$and ${\cal Z}$ is the sample partition function for a given configuration of the disorder fields. For the current model, we have \cite{FAR}
\begin{equation}
 {\cal Z} =\int {\cal D}u\,\rme^{-\beta {\cal H}[u; \{e_i\}]}, 
\label{eq:sample_Z}
\end{equation}
where ${\cal H}={\cal H}_{\mathrm{b}}+{\cal H}_{\mathrm{s}}$ and  $\beta=1/(\kBT)$.  To carry out the averages  involved here,  we use the replica trick~\cite{dotsenko1,dotsenko2} by considering $k$ replicated fields $u_a$, with $a=1,\cdots, k$, to generate the replicated partition function ${\cal Z}^k$ and then compute the replicated free energy as \cite{partial}
\begin{equation}
{\cal F}^{(k)}=-\kB T\frac{\ln \langle {\cal Z}^k \rangle}{k}. 
\label{eq:rep_free}
\end{equation}
The quenched and annealed cases then follow for $k\rightarrow 0$ and $k=1$, respectively. The quenched free energy can also be expressed as 
${\cal F}_Q=-\kB T\partial_k \langle {\cal Z}^{k} \rangle_{k\rightarrow 0}$.  The special case of a nondisordered sm-A~\cite{FAR} follows from our formalism, when the disorder variance is set to zero. 

\section{Formalism}
\label{sec:formalism}

\subsection{Replicated partition function}

The disorder-averaged replicated partition function in Eq. \eqref{eq:rep_free} follows by carrying out the standard Gaussian averages over the disorder fields over the sample partition function \eqref{eq:sample_Z}, yielding  
\begin{equation}
\langle {\cal Z}^{k} \rangle=\int \Big(\prod_{a=1}^{k} {\cal D}u_{a}\Big)\,\rme^{-\beta {\cal H}^{(k)}[\{u_{a}\}]}, 
\label{gfnchjkw}
\end{equation}
where ${\cal H}^{(k)}={\cal H}_{\mathrm{b}}^{(k)}+{\cal H}_{\mathrm{s}}^{(k)}$ with the definitions 
 \begin{eqnarray}
\!\!\!\!\!{\cal H}_{\mathrm{b}}^{(k)}=\frac{K}{2}\sum_{a=1}^{k}\int_{V} \rmd{\mathbf r} \left[\left(\partial_{x}^2 u_a+\partial_{y}^2 u_a\right)^2+{B\over K}(\partial_{z}u_a)^2\right],
\label{bulkK}
\end{eqnarray}
\begin{eqnarray}
&&{\cal H}_{\mathrm{s}}^{(k)}=
\frac{W}{2}\sum_{a,b=1}^{k}\int_{\partial V} \rmd{\mathbf r}_{\!_\perp}\rmd{\mathbf r^{\prime}}_{\!_\perp}\nonumber\\
&& \bigg[(\partial_{x}u_{a})\Big(\delta({\mathbf r}_{\!_\perp}-{\mathbf r}_{\!_\perp}^{\prime})\delta_{ab}-\beta W c({\mathbf r}_{\!_\perp}-{{\mathbf r}_{\!_\perp}^{\prime}})E_{ab}\Big)(\partial_{x^{\prime}}u_{b})+\nonumber\\
&&(\partial_{y}u_{a})\Big(\delta({\mathbf r}_{\!_\perp}-{\mathbf r}_{\!_\perp}^{\prime})\delta_{ab}-\beta W c({\mathbf r}_{\!_\perp}-{{\mathbf r}_{\!_\perp}^{\prime}})E_{ab}\Big)(\partial_{y^{\prime}}u_{b})\bigg],
\label{surfaceK}
\end{eqnarray}
where $\delta_{ab}$ is the Kronecker delta and  $E_{ab}=1$; these are the elements of the $k\times k$ identity  matrix, ${\mathbb I}$, and the $k\times k$ matrix of ones, ${\mathbb E}$, respectively. 

\subsection{Fourier diagonalization}

Assuming translational invariance
in the $xz$-plane, we can make use of the Fourier transformation by defining $u({\mathbf r}_{\!_\perp},y)=\sum_{{\mathbf p}_{\!_\perp}}u({\mathbf p}_{\!_\perp},\!y)\,\rme^{\icomplex {\mathbf p}_{\!_\perp}\!\cdot{\mathbf r}_{\!_\perp} }$, where ${\mathbf p}_{\!_\perp}=(p_x, p_z)$ is the corresponding in-plane wavevector. The replicated Hamiltonian is then diagonalized over the subspace defined by the in-plane modes and can be written as 
\begin{equation}
{\cal H}^{(k)}=\sum_{{\mathbf p}_{\!_\perp}}(h_{\mathrm{b}}^{(k)}+h_{\mathrm{s}}^{(k)}),
\label{H_rep}
\end{equation}
where the bulk and surface terms follow from Eqs.~\eqref{bulkK} and 
\eqref{surfaceK} as 
\begin{widetext}
\begin{eqnarray}
\label{ham_b}
&&h_{\mathrm{b}}^{(k)}\big[\{u_{a}({\mathbf p}_{\!_\perp},y)\}\big]=\frac{KA}{2}\sum_{a=1}^k\!\int_{0}^{d}\! \rmd y\, \Big\{ |\ddot u_{a}({\mathbf p}_{\!_\perp},y)|^2 + 2 p_x^2 |\dot u_{a}({\mathbf p}_{\!_\perp},y)|^2\,+\, (p_x^4 + \lambda^{-2} p_z^2)|u_{a}({\mathbf p}_{\!_\perp},y)|^2\Big\},
\\
\label{ham_s}
&&h_{\mathrm{s}}^{(k)}\big(\{u_{a}({\mathbf p}_{\!_\perp},d)\}\big) \,= \frac{WA}{2}\sum_{a,b=1}^k \Big\{ \dot u_{a}({\mathbf p}_{\!_\perp},d)\Big(\delta_{ab}-\beta W c({\mathbf p}_{\!_\perp})E_{ab}\Big)\dot u_{b}^{\ast}({\mathbf p}_{\!_\perp}, d)+\nonumber\\
&&\qquad\qquad + \, p_x^2 u_{a}({\mathbf p}_{\!_\perp},d)\Big(\delta_{ab}-\beta W c({\mathbf p}_{\!_\perp})E_{ab}\Big) u_{b}^{\ast}({\mathbf p}_{\!_\perp},d)\Big\} 
-\frac{KA}{2} \sum_{a=1}^k p_x^2 \Big\{u_{a}({\mathbf p}_{\!_\perp},d) \dot u_{a}^{\ast}({\mathbf p}_{\!_\perp},d)+c.c.\Big\},
\end{eqnarray}
\end{widetext}
where a surface contribution resulting form an integration by part of the bulk terms, Eq. \eqref{ham_b}, has implicitly been calculated and the result is included as the last term of Eq. \eqref{ham_s}.  
In the above relations, $\lambda=\sqrt{K/B}$ is the {\em penetration length}~\cite{de-gennes}, $A$ is the surface area of each plate,
$u_{a}^{*}({\mathbf p}_{\!_\perp},y)=u_{a}(-{\mathbf p}_{\!_\perp},y)$, $c({\mathbf p}_{\!_\perp})$ is the Fourier transform of $c({\mathbf r}_{\!_\perp}-\mathbf r^{\prime}_{\!_\perp})$, and the $y$-derivative is denoted as ${\dot u}_{a}({\mathbf p}_{\!_\perp},y)\equiv \partial_{y}u_{a}({\mathbf p}_{\!_\perp},y)$.

Using Eqs.~(\ref{H_rep})-(\ref{ham_s}) and the path-integral methods discussed in Refs. \cite{kleinert,handbook}, we find (see Ref. \cite{FAR} for complementary  details not reiterated here) 
\begin{widetext}
	\begin{eqnarray}
&&	 \langle {\cal Z}^{k} \rangle= \prod_{{\mathbf p}_{\!_\perp}} \int  
	\bigg\{\prod_{a=1}^{k}\rmd(u_{a}({\mathbf p}_{\!_\perp},d))\,\rmd(\dot u_{a}({\mathbf p}_{\!_\perp},d))\bigg\}  \,  {\cal G}\Big( 0, 0; \dot u_{a}({\mathbf p}_{\!_\perp},d), u_{a}({\mathbf p}_{\!_\perp},d)\Big) \rme^{- \beta h_{\mathrm{s}}^{(k)}(\{u_{a}({\mathbf p}_{\!_\perp},d)\})},
	\label{partition_fuction}\\
&& {\cal G}\Big(\!0, 0; \dot u_{a}({\mathbf p}_{\!_\perp},d), u_{a}({\mathbf p}_{\!_\perp},d)\!\Big) \!=\!\!\int \!  \Big(\prod_{a=1}^{k} {\cal D}u_{a}({\mathbf p}_{\!_\perp},y) \Big) \,\rme^{- \beta h_{\mathrm{b}}^{(k)}[\{u_{a}({\mathbf p}_{\!_\perp},y)\}]}. 
	\end{eqnarray}
\end{widetext}
The first two arguments of the Green's function, $ {\cal G}\Big(\!0, 0; \dot u_{a}({\mathbf p}_{\!_\perp},d), u_{a}({\mathbf p}_{\!_\perp},d)\!\Big) $, indicate the strong anchoring at the substrate $y = 0$, while its last two arguments give the values of the fields $\dot u_a$ and $u_a$ at $y=d$. This interim result can be simplified further by integrating over the bulk displacement fields $u_{a}({\mathbf p}_{\!_\perp},y)$, subjected to the given boundary conditions,
yielding the replicated partition function only in terms of integrals over the surface fields $\{u_{a}({\mathbf p}_{\!_\perp},d),\dot u_{a}({\mathbf p}_{\!_\perp},d)\}$ weighted by an effective surface Hamiltonian $h_{\mathrm{s}}^{\mathrm{eff}}[\{u_{a}({\mathbf p}_{\!_\perp},d), \dot u_a ({{\mathbf p}_{\!_\perp}}\!,d)\}]$; i.e., 
\begin{widetext}
\begin{eqnarray}
&&\langle {\cal Z}^{k} \rangle=  \prod_{{\mathbf p}_{\!_\perp}}{\cal A}^{k}(\omega_1,\omega_2,d)\!\int \! \bigg\{\prod_{a=1}^{k}\rmd(u_{a}({\mathbf p}_{\!_\perp},d))\,\rmd(\dot u_{a}({\mathbf p}_{\!_\perp},d))\bigg\}\,\rme^{- \beta h_{\mathrm{s}}^{\mathrm{eff}}[\{u_{a}({\mathbf p}_{\!_\perp},d),\dot u_{a}({\mathbf p}_{\!_\perp},d)\}]}, 
 \label{rep_Z}
 \\
&&h_{\mathrm{s}}^{\mathrm{eff}}=\sum_{a,b=1}^k\bigg[m_{11}^{(k)}  u_{a}({\mathbf p}_{\!_\perp},d) u_{b}^{\ast}({\mathbf p}_{\!_\perp},d) 
+\,m_{13}^{(k)} \big(\dot u_{a}({\mathbf p}_{\!_\perp},d) u_{b}^{\ast}({\mathbf p}_{\!_\perp},d) +c.c.\big)
+\,m_{33}^{(k)} \dot u_{a}({\mathbf p}_{\!_\perp},d)\dot u_{b}^{\ast}({\mathbf p}_{\!_\perp},d)\bigg],
\label{eq:h_s_eff}
\end{eqnarray}
\end{widetext}
where we find
\begin{eqnarray}
&&m_{11}^{(k)}=m_{11}\delta_{ab}-{\beta W c({\mathbf p}_{\!_\perp})p_x^2\over 2\ell}E_{ab}, 
\label{m11}
\\
&&m_{33}^{(k)}=m_{33}\delta_{ab}-{\beta W c({\mathbf p}_{\!_\perp})\over 2\ell}E_{ab}, 
\label{m33}
\\
\label{m13}
&&m_{13}^{(k)}=m_{13}\delta_{ab}, 
\end{eqnarray}
with the definitions 
\begin{eqnarray}
&&m_{11}=\frac{\omega_1 \omega_2}{2{\mathcal B}} (\omega_{1}^2-\omega_{2}^2)(\omega_{1} s_{1}c_{2}-\omega_{2} c_{1} s_{2})
+{p_x^2\over 2\ell}, 
\\
&&m_{33}={1\over 2{\mathcal B}}(\omega_{1}^2-\omega_{2}^2)(\omega_{1} s_{2}c_{1}-\omega_{2} c_{2} s_{1})
+{1\over 2\ell},
\\
&&m_{13}=-\frac{\omega_1 \omega_2}{2{\mathcal B}} [(\omega_{1}^2+\omega_{2}^2)(c_1 c_2 -1)-2\omega_1 \omega_2 s_1 s_2]
-{p_x^2\over 2}, \nonumber\\
\label{defmik}
\end{eqnarray}
and
\begin{eqnarray}
\label{omega0}
&&\omega_1 ^2+\omega_2 ^2=2 p_x ^2,\\
&&\omega_1 ^2\omega_2 ^2=p_x ^4+\lambda^{-2}p_z^2. 
\label{omega}
\end{eqnarray}
Here, $\ell=K/W$ is the {\em extrapolation length} and the factor ${\cal A}$
is  given by~\cite{kleinert} 
\begin{equation}
{\cal A}(\omega_1, \omega_2, d)={1\over 2\pi}{\sqrt{\omega_1 \omega_2}|\omega_1^2-\omega_2^2|\over \sqrt{{\mathcal B}}},
\label{F_beta}
\end{equation}
with 
\begin{equation} 
{\mathcal B}= 
(\omega_1^2+\omega_2^2)s_1 s_2-2\omega_1 \omega_2 (c_1 c_2 -1)
,
\label{detM}
\end{equation}
and the shorthand notations $c_i\equiv\cosh (\omega_i d) $ and $s_i\equiv\sinh (\omega_i d)$ for $i=1,2$~\cite{kleinert}.

\subsection{The disorder-averaged free energy}
\label{free_energy}

The final solution to the path-integral representation of the replicated partition function,  Eq.~(\ref{rep_Z}), is 
\begin{equation}
 \langle {\cal Z}^{k} \rangle= \prod_{{\mathbf p}_{\!_\perp}}{\cal A}^{k}\rme^{-{1\over 2}\ln \det {\mathbb M}^{(k)}}= \prod_{{\mathbf p}_{\!_\perp}}\rme^{-{k\over 2}\ln |{\mathcal B}|-{1\over 2} \ln \det {\mathbb M}^{(k)}},
 \label{Z_m}
\end{equation}
where we have omitted irrelevant prefactors coming from ${\cal A}$, and defined the matrix ${\mathbb M}^{(k)}$ as 
\begin{equation}
{\mathbb M}^{(k)}=\begin{pmatrix}
	m_{11}^{(k)} & -m_{13}^{(k)} \\
	\\
	-m_{13}^{(k)} & m_{33}^{(k)}
\end{pmatrix},
\label{eq:M_k}
\end{equation}
whose elements, Eqs.~(\ref{m11}), (\ref{m33}) and (\ref{m13}), are each a $k\times k$ matrix. In other words, ${\mathbb M}^{(k)}$ can be expressed using the Kronecker product  as 
${\mathbb M}^{(k)} = {\mathbb M}\otimes{\mathbb I} -{(\beta W c({\mathbf p}_{\!_\perp})/ 2)}\,{\mathbb P}\otimes{\mathbb E}$, where the $2\times 2$ matrices ${\mathbb M}$ and ${\mathbb P}$ are defined as
\begin{equation}
{\mathbb M}
=\begin{pmatrix}
m_{11} & -m_{13} \\
\\
-m_{13} & m_{33}
\end{pmatrix},
\quad
{\mathbb P}
=\begin{pmatrix}
p_x^2/\ell & 0 \\
\\
0 & 1/\ell
\end{pmatrix}.
\end{equation}

Due to its particular form, 
the determinant of ${\mathbb M}^{(k)}$ can then be obtained using known matrix identities \cite{partial}, giving  
\begin{equation}
\ln \det {\mathbb M}^{(k)}= k\ln \det {\mathbb M} + 
 \ln \det \Big({\mathbb I}
-k {\beta W c({\mathbf p}_{\!_\perp})\over 2}{\mathbb P}\,{\mathbb M}^{-1}\Big).
\end{equation}

The replicated partition function can be written explicitly using the above relation. The thermodynamic free energy of the system, ${\cal F}$,  can then be derived in the two limiting cases of interest using Eqs. \eqref{eq:rep_free} and \eqref{Z_m}. Hence,  in the case of {\em annealed disorder} ($k=1$), we find  
\begin{equation}
{\cal F}_A={\kB T\over 2}\sum_{{\mathbf p}_{\!_\perp}}\ln \left|{\mathcal B}\det \left({\mathbb M}- {\beta W c({\mathbf p}_{\!_\perp})\over 2}{\mathbb P}\right)\right|, 
\label{eq:free_ann}
\end{equation}
and in the case of {\em quenched disorder} ($k\rightarrow 0$), 
\begin{equation}
{\cal F}_Q={\kB T\over 2}\sum_{{\mathbf p}_{\!_\perp}}\ln \left|{\mathcal B}\det {\mathbb M}\right| - {W\over 4} \sum_{{\mathbf p}_{\!_\perp}}c({\mathbf p}_{\!_\perp})\,\tr \, ({\mathbb P}\,{\mathbb M}^{-1}). 
\label{eq:free_qnch}
\end{equation}

The contribution due to quenched disorder is thus completely decoupled from the standard thermal pseudo-Casimir free energy \cite{FAR}  and contributes a term proportional to the disorder variance, in agreement with the general paradigm discussed in the context of charge disorder \cite{sndhp,ddnp}. The free energy in the quenched limit can be rewritten explicitly as 
\begin{eqnarray}
\label{G}
&& {\cal F}_Q={\kB T\over 2}\sum_{{\mathbf p}_{\!_\perp}}\ln \left|{\mathcal B}\left(m_{11}m_{33}-m_{13}^2\right)\right| 
\\
&&\qquad\qquad\qquad\quad-{W \over 4\ell}\sum_{{\mathbf p}_{\!_\perp}} c({\mathbf p}_{\!_\perp})\,{p_x^2 m_{33}
	+m_{11}\over 
	m_{11}m_{33}-m_{13}^2}. \nonumber
\end{eqnarray}  

\section{Results}
\label{sec:force}

\subsection{Quenched disorder}
\label{ec:force_ana}

We proceed by determining the final forms of the terms contributing to the  free energy ${\cal F}$ in Eq.~(\ref{G}) in the current slab geometry. We consider the quenched disorder case in this section and return to the case of annealed disorder later in Section \ref{ec:force_ana_an}. 

We denote the argument of the logarithm in the disorder-free contribution, which is the first term in Eqs.~\eqref{eq:free_qnch} and \eqref{G}, by 
 \begin{equation}
G_0({\mathbf p}_{\!_\perp})\equiv {\mathcal B}\det {\mathbb M}. 
\end{equation}
This quantity can be calculated explicitly as  
\begin{eqnarray}
2 G_0({\mathbf p}_{\!_\perp})\,=A&&\,+\,B\cosh (2\alpha d)+C \sinh (2\alpha d)\nonumber\\
&&\,+\,D\cos (2\gamma d)+E\sin (2\gamma d),
\label{green}
\end{eqnarray}
where $\alpha$ and $\gamma$ give the solutions of Eqs.~(\ref{omega0}) and (\ref{omega}) as $\omega_1=\alpha-\icomplex\gamma$ and $\omega_2=\alpha+\icomplex\gamma$, where
\begin{eqnarray}
&&\alpha({\mathbf p}_{\!_\perp}) =\frac{1}{\sqrt{2}}\left(p_{x}^2+\sqrt{p_{x}^4+{\lambda^{-2}p_z^2}}\right)^{1/2},
\label{alpha}
\\
&&\gamma({\mathbf p}_{\!_\perp}) =\frac{1}{\sqrt{2}}\left(-p_{x}^2+\sqrt{p_{x}^4+{\lambda^{-2}p_z^2}}\right)^{1/2}, 
\label{gamma}
\end{eqnarray}
and 
\begin{eqnarray}
&&A({\mathbf p}_{\!_\perp})=-\sqrt{p_x^4 + \lambda^{-2}p_z^2}\left(\lambda^{-2}p_z^2 - \ell^{-2}p_x^2\right),
\label{coefficients1} \\
&&B({\mathbf p}_{\!_\perp})=-\gamma^2 \left(\lambda^{-2} p_z^2 +\,\ell^{-2} p_x^2\right),
\label{coefficients2} \\
&&C({\mathbf p}_{\!_\perp})=-\alpha\,\lambda^{-2}p_z^2 \ell^{-1} 
\label{coefficients3} , \\
&&D({\mathbf p}_{\!_\perp})=-\alpha^2 \left(\lambda^{-2} p_z^2 +\ell^{-2} p_x^2\right),
\\
&&E({\mathbf p}_{\!_\perp})=-\gamma\,\lambda^{-2}p_z^2 \ell^{-1}. 
\label{coefficients}
\end{eqnarray}
In the disorder-induced component, which is the second term in Eqs.~\eqref{eq:free_qnch} and \eqref{G}, we define
\begin{equation}
G_{d}({\mathbf p}_{\!_\perp}) \equiv \ell^{-1}{\mathcal B}\left(p_x^2 m_{33} +m_{11}\right), 
\end{equation}
which can then be written as 
\begin{eqnarray}
G_{d}({\mathbf p}_{\!_\perp})=A_{d}&&\,+\,B_{d}\cosh (2\alpha d)+C_{d} \sinh (2\alpha d)\nonumber\\
&&\,+\,D_{d}\cos (2\gamma d)+E_{d}\sin (2\gamma d),
\label{green_d}
\end{eqnarray}
where
\begin{eqnarray}
&&A_{d}({\mathbf p}_{\!_\perp})=2 \ell^{-2}p_x ^2\sqrt{p_x^4 + \lambda^{-2}p_z^2}, \\
&&B_{d}({\mathbf p}_{\!_\perp})=-2 p_x ^2\ell^{-2} \gamma^2,\\
&&C_{d}({\mathbf p}_{\!_\perp})=-\lambda^{-2}p_z^2 \ell^{-1} \alpha, \\
&&D_{d}({\mathbf p}_{\!_\perp})=-2 p_x ^2 \ell^{-2} \alpha^2, \nonumber\\
&&E_{d}({\mathbf p}_{\!_\perp})=-\lambda^{-2}p_z^2 \ell^{-1}\gamma. 
\label{coefficients_d}
\end{eqnarray}
The quenched free energy can thus be rewritten as 
 \begin{equation}
  {\cal F}(d)={\cal F}_0 (d) + {\cal F}_{dis} (d),
  \label{bcfgjykab}
 \end{equation}
with the disorder-free and disorder-induced components being, respectively, 
\begin{eqnarray}
&&{\cal F}_0 (d)={\kB T\over 2}\sum_{{\mathbf p}_{\!_\perp}}\ln |G_0({\mathbf p}_{\!_\perp})|, 
\\
&&{\cal F}_{dis} (d)=-{W\over 4} \sum_{{\mathbf p}_{\!_\perp}} ~c({\mathbf p}_{\!_\perp}){G_{d} ({\mathbf p}_{\!_\perp})\over G_0({\mathbf p}_{\!_\perp})}.
\end{eqnarray}
The disorder-free term has been studied extensively in Ref. \cite{FAR}, following which the disorder-induced term can be written in the continuum limit (assuming large areas, $A$, for each of the  bounding substrates) as
\begin{equation}
{\cal F}_{dis} (d)=-{WA\over 4\pi^2}\! \int_{0}^\infty\!\! \rmd p_x \!\int_{0}^{\pi/a_0}\!\!\rmd p_z ~c(p_x ,p_z){G_{d} (p_x ,p_z)\over G_0(p_x ,p_z)},
\label{eq:Fdis_2}
\end{equation}
where we have taken the $p_z$ integral in the first Brillouin zone $-\pi/a_0\leq p_z<\pi/a_0$. The above free energy also contains the bulk and the surface self-energy parts, which can be derived in the limit of $d \rightarrow \infty$ in the following forms
\begin{eqnarray}
&&{\cal F}_0 \left(d \rightarrow \infty\right)={\kB T}\, d\, \sum_{{\mathbf p}_{\!_\perp}}\alpha({\mathbf p}_{\!_\perp}) + {\textrm{const.}},
\label{bdxjkqw}
\\
&&{\cal F}_{dis} \left(d \rightarrow \infty\right)=-{W\over 4} \sum_{{\mathbf p}_{\!_\perp}} c({\mathbf p}_{\!_\perp}){{B({\mathbf p}_{\!_\perp}) + C({\mathbf p}_{\!_\perp})}\over{B_{d}({\mathbf p}_{\!_\perp})+C_{d}({\mathbf p}_{\!_\perp})}},\nonumber\\
~
\end{eqnarray}
with the bulk part being proportional to the volume, i.e., $A\cdot d$, and the surface self-energy proportional to the substrate anchoring 
energy, $A\cdot W$, in the continuum representation. These two contributions are subtracted from the total free energy Eq. \eqref{bcfgjykab} in order to get the proper fluctuation-induced interaction free energy between the two bounding substrates.

While the relation \eqref{eq:Fdis_2} can now be evaluated conveniently using numerical methods, one may note an interesting analytical aspect of the problem; that the disorder-induced contribution, ${\cal F}_{dis}$, does not lead to any additional interaction forces beyond those given by the disorder-free term, ${\cal F}_0$, neither in the limit of weak ($W\rightarrow 0$, or $\lambda/\ell\rightarrow 0$) nor in the limit of strong ($W\rightarrow\infty$, or $\lambda/\ell\rightarrow \infty$) anchoring. While in the former limit, ${\cal F}_{dis}$ vanishes, in the latter limit, ${\cal F}_{dis}=-W \sum_{{\mathbf p}_{\!_\perp}} c({\mathbf p}_{\!_\perp})$, contributing  only a surface free energy. The disorder-free term itself will be different in the two limits as we find  
\begin{equation}
G_0({\mathbf p}_{\!_\perp})\big|_{W\rightarrow 0}=-\lambda^{-2}p_z ^2 \big[\gamma^2 \cosh ^2 (\alpha d)+\alpha^2 \cos^2 (\gamma d)\big],
\end{equation}  
in the weak anchoring, and 
\begin{equation}
G_0({\mathbf p}_{\!_\perp})\big|_{W\rightarrow \infty}=-\ell^{-2}p_x ^2 \big[\gamma^2 \sinh ^2 (\alpha d)-\alpha^2 \sin^2 (\gamma d)\big], 
\end{equation}  
in the strong anchoring limit. 

\begin{figure}[t!]
	\begin{center}
		\includegraphics[width=7.7cm]{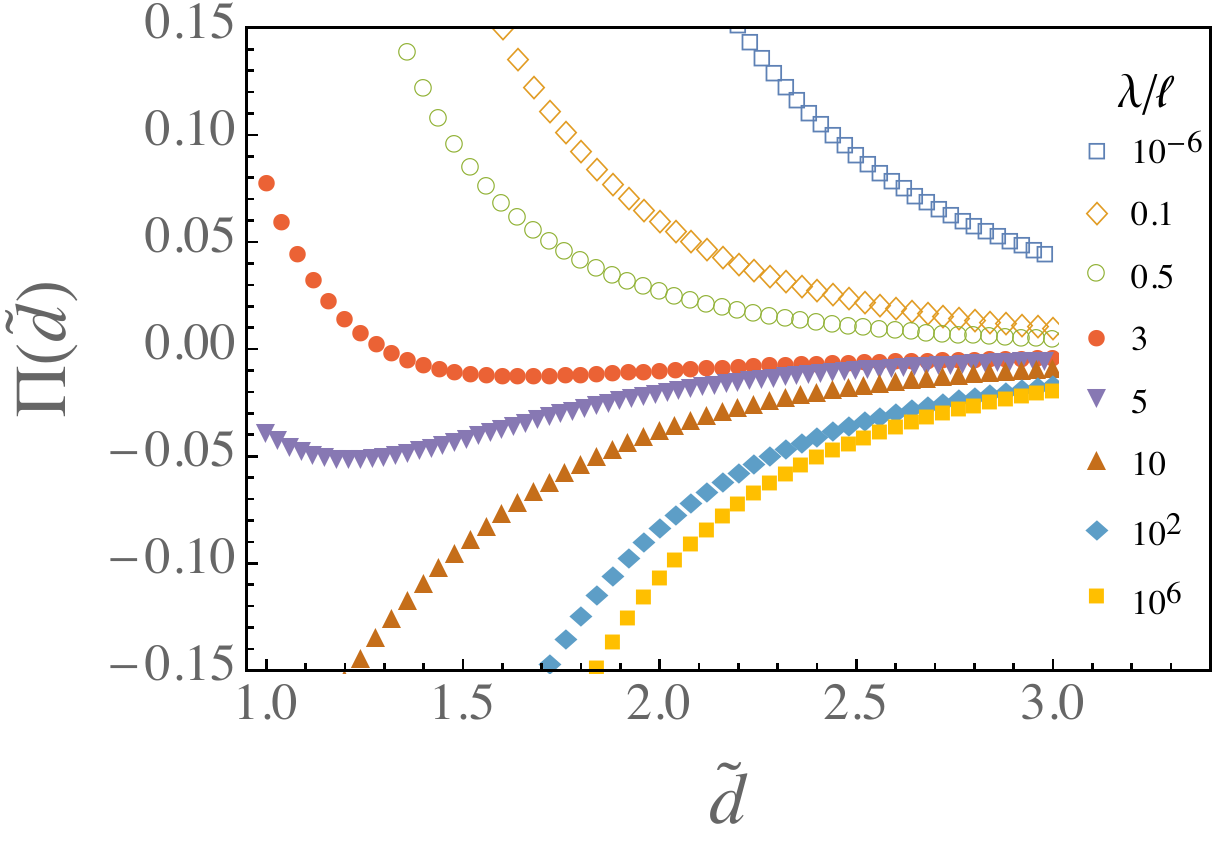}
		\caption{Rescaled interaction pressure ${\Pi}(\tilde d)$ as a function of the rescaled separation $\tilde d=d/\lambda$ in the disorder-free case ($c_0=0$) for  $\lambda/a_0=2$ and $\lambda/\ell=10^{6}, 100, 10, 5, 3, 0.5, 0.1$, and $10^{-6}$ (bottom to top, limiting cases of strong to weak anchoring). 
		}
		\label{fig:fig2}
	\end{center}
\end{figure}

The fluctuation-induced force between the bounding surfaces of the sm-A film can be calculated by differentiating the free energy of the system with respect to the inter-surface distance, $d$,  as 
\begin{equation} 
\label{eq:force}
f(d) = -\frac{\partial {\cal F}(d)}{\partial d}. 
\end{equation}
The corresponding pressure acting on the bounding surfaces can be represented in dimensionless form as a rescaled pressure $\Pi(\tilde d)$ as a function of the rescaled inter-surface separation $\tilde d=d/\lambda$  
\begin{equation} 
\label{eq:pressure}
\Pi(\tilde d)=  \frac{\beta \lambda^3}{A}f(\lambda \tilde d). 
\end{equation}
This thermal  pseudo-Casimir interaction pressure  is calculated and analyzed after the bulk pressure part, corresponding to ${\cal F}_0 (d \rightarrow \infty)$ in Eq. \eqref{bdxjkqw}, is  subtracted.

Figure \ref{fig:fig2} presents the rescaled pressure as a function of the rescaled inter-surface separation for various strengths of the surface anchoring on the substrate located at $y=d$, which is  measured by the ratio $\lambda/\ell$. We vary $\lambda/\ell$ in a wide range of small to large values, from top to bottom, to capture the limiting cases of $\lambda/\ell\rightarrow 0$ and $\infty$. As the strength of the anchoring is reduced (smaller $\lambda/\ell$), the magnitude of the attractive force between the bounding substrates of the sm-A film diminishes. This mirrors the fact that the boundary conditions become increasingly non-similar (asymmetric) on the two substrates (given that one of the substrates is always kept at strong anchoring condition; see Section \ref{sec:model}), where repulsive interaction forces are expected. The repulsive pseudo-Casimir force is therefore purely a consequence of the asymmetry of the boundary conditions just as in the case of the van der Waals interactions with asymmetric (dissimilar)  boundary conditions \cite{Parsegian}. Similar behavior is found in the case of nematic  films~\cite{LC3a,LC3b}. 

A closer inspection of the force behavior across the intermediate ranges of the anchoring strengths ($\lambda/\ell\sim 1$) reveals non-monotonic interaction pressure profiles with a minimum found at intermediate values of $\tilde d$. 
Clearly, this kind of behavior is an effect of the competing  boundary conditions on the two bounding substrates, or equivalently, of the different degrees of the boundary condition asymmetry in the system. The effect of this asymmetry is evidently more pronounced at intermediate separations, leading to a change in sign when compared with the asymptotic large- and small-separation regimes.  

\begin{figure}[t!]
	\begin{center}
		\includegraphics[width=7.7cm]{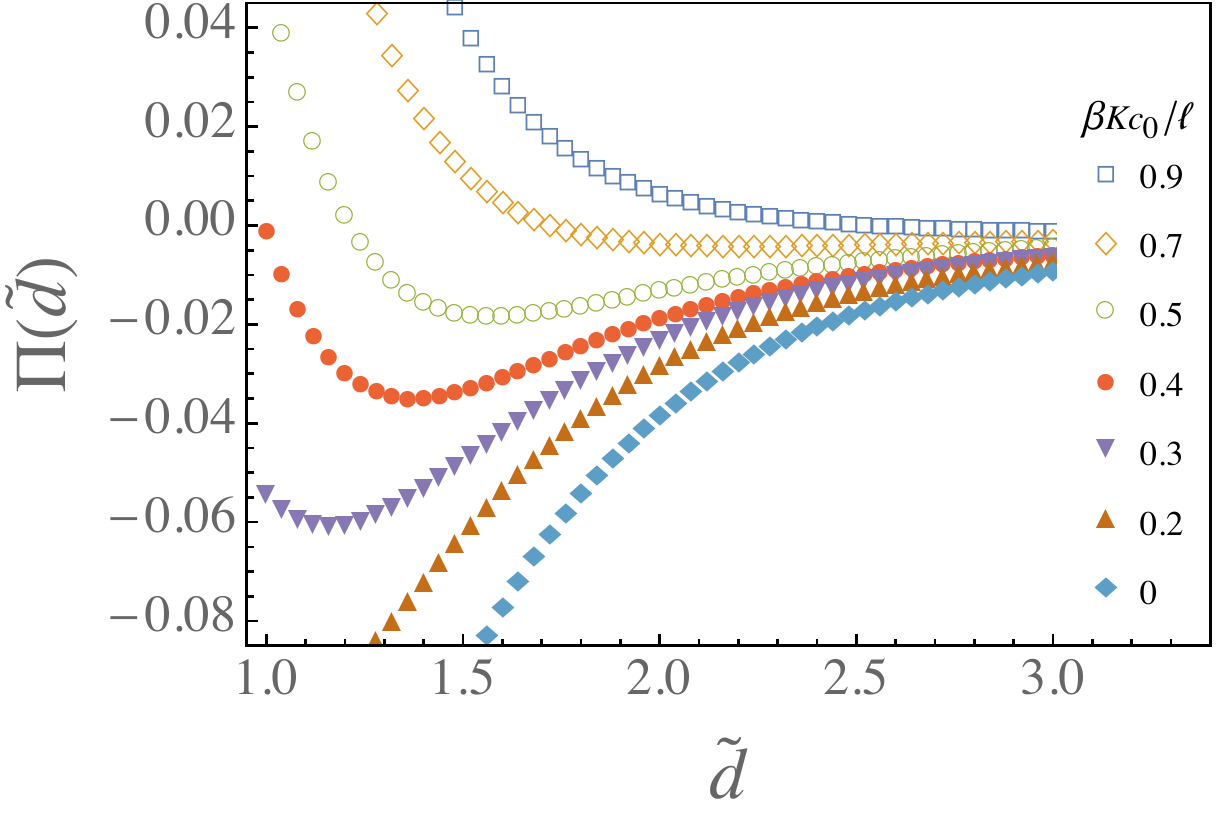}
		\caption{Rescaled interaction pressure ${\Pi}(\tilde d)$ as a function of the rescaled separation $\tilde d=d/\lambda$ for  $\lambda/a_0=2$,  $\lambda/\ell =10$, and  uncorrelated quenched surface disorder of rescaled variances  $\beta K c_0/\ell=0, 0.2, 0.3, 0.4, 0.5, 0.7$ and $0.9$ (bottom to top).
		}
		\label{fig:fig3}
	\end{center}
\end{figure}

Figure \ref{fig:fig3}, on the other hand, shows the effects of a finite degree of quenched disorder as quantified by its variance. The surface disorder is chosen here to be uncorrelated and have a constant variance, i.e., $c({\mathbf p}_{\!_\perp}) = c_0$, and the anchoring strength (on the substrate at $y=d$) is fixed using $\lambda/\ell =10$, in which case, we find an attractive interaction pressure in the absence of disorder (blue diamonds, bottom-most data set).  Increasing the strength of the disorder through its variance, reduces the magnitude of the attractive pressure and even changes the sign of the interaction pressure into a positive (repulsive) one, displaying also an intermediate regime with non-monotonic pressure profiles. The quenched disorder thus acts as an additional source of boundary asymmetry, strengthening the effects of dissimilar bounding substrates.

\begin{figure}[t!]
	\begin{center}
		\includegraphics[width=7.7cm]{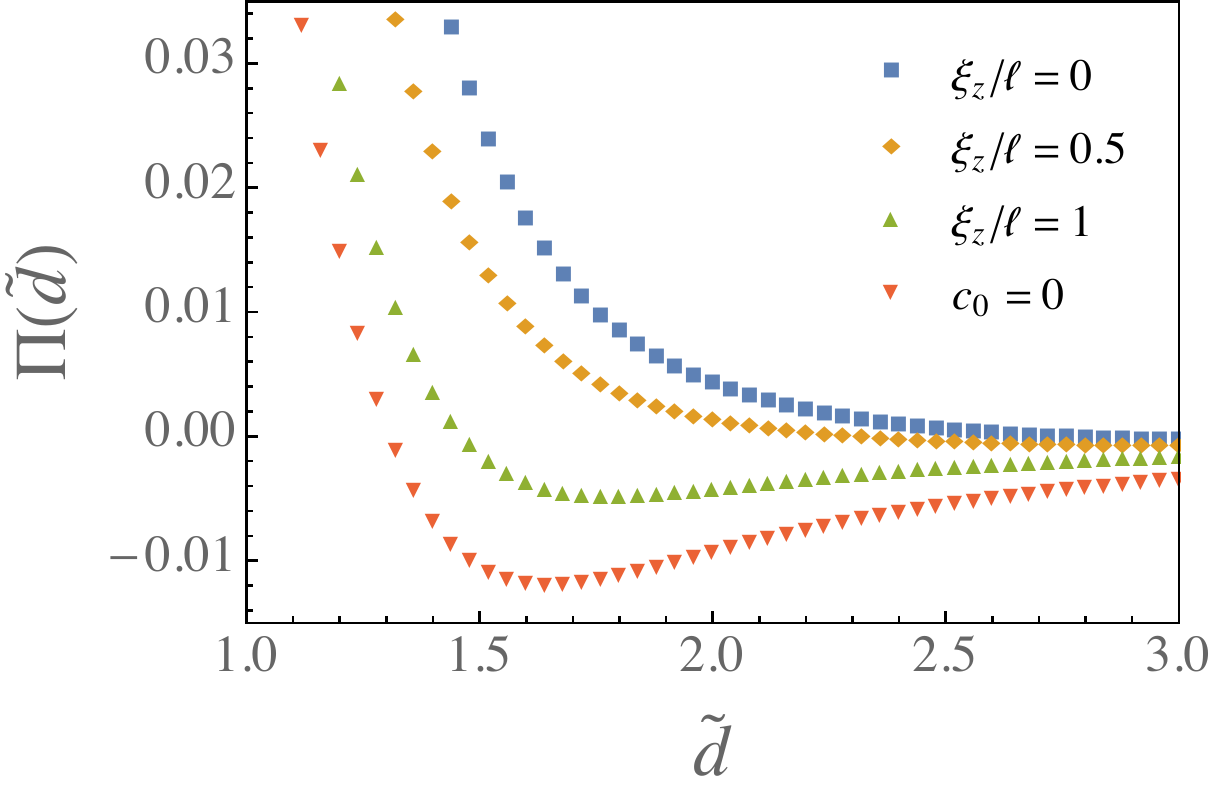}
		\caption{Rescaled interaction pressure ${\Pi}(\tilde d)$ as a function of the rescaled separation $\tilde d=d/\lambda$ for fixed $\lambda/a_0=2$,  $\lambda/\ell =3$, and with a correlated quenched surface disorder (see Eq. \eqref{eq:corr}) of rescaled variance  $\beta K c_0/\ell=0.3$ and rescaled correlation length of $\xi_z/\lambda=0$ (uncorrelated case), 0.5 and 1 from top to bottom. The disorder-free results ($c_0=0$) are shown by red triangle-downs (bottom-most data set). 
		} 
		\label{fig:fig3p}
	\end{center}
\end{figure}

To model a spatially correlated surface disorder, we may use the fact that the in-plane correlation length of the random surface pinning may be anisotropic and given by the relation 
 $\xi_z\simeq \lambda^{-1}\xi_x^2$, where $\xi_{z}$ and $\xi_{x}$ are the correlation lengths of the molecular-axis disorder along the $z$ and $x$ directions of our sm-A geometry, respectively~\cite{radzihovsky3}. Accordingly, we choose the form of the in-plane disorder correlation function as 
 \begin{equation}
 c(p_x ,p_z )=\frac{c_0}{\xi_z^4(p_x^4 +\lambda^{-2}p_z^2)+1}.
 \label{eq:corr}
 \end{equation}
As one notes by increasing the correlation length $\xi_z$, the role of disorder diminishes. Figure~\ref{fig:fig3p} exemplifies this behavior for the model parameter values $\lambda/a_0=2$, $\lambda/\ell=3$, $\beta K c_0/\ell=0.3$, and $\xi_z/\lambda=0, 0.5$ and 1 (from top to bottom). For comparison, we also plot the interaction pressure in the disorder-free case  ($c_0=0$)  in Fig. \ref{fig:fig3p} (red triangle-downs),
which consistently falls below the disordered cases shown. 

\subsection{Annealed disorder}
\label{ec:force_ana_an}

 We now turn to the case of annealed disorder in which case the interaction free energy \eqref{eq:free_ann} can be written explicitly as    
\begin{equation}
{\cal F}_A (d)={\kB TA\over 2\pi^2} \int_{0}^\infty \rmd p_x \!\int_{0}^{\pi/a_0}\rmd p_z ~\ln |G_A(p_x ,p_z)|, 
\end{equation} 
 where $G_A$ turns out to be of the exact same form as $G_0$ in Section \ref{ec:force_ana}, except that here one has to make the replacement  $\ell^{-1}\rightarrow \ell_{eff}^{-1}\equiv \ell^{-1}(1-\beta W c({\mathbf p}_{\!_\perp}))$. In other words, the extrapolation length is renormalized to an effective one due to the annealed disorder,  reflecting the non-additive modification this type of disorder makes in the fluctuation-induced force between the bounding substrates. To ensure that the equilibrium order-parameter profile of the sm-A film remains unchanged, the factor $(1-\beta W c({\mathbf p}_{\!_\perp}))$ should be kept positive for all ${\mathbf p}_{\!_\perp}$-modes.  
 
\begin{figure}[t!]
 	\begin{center}
		\includegraphics[width=7.7cm]{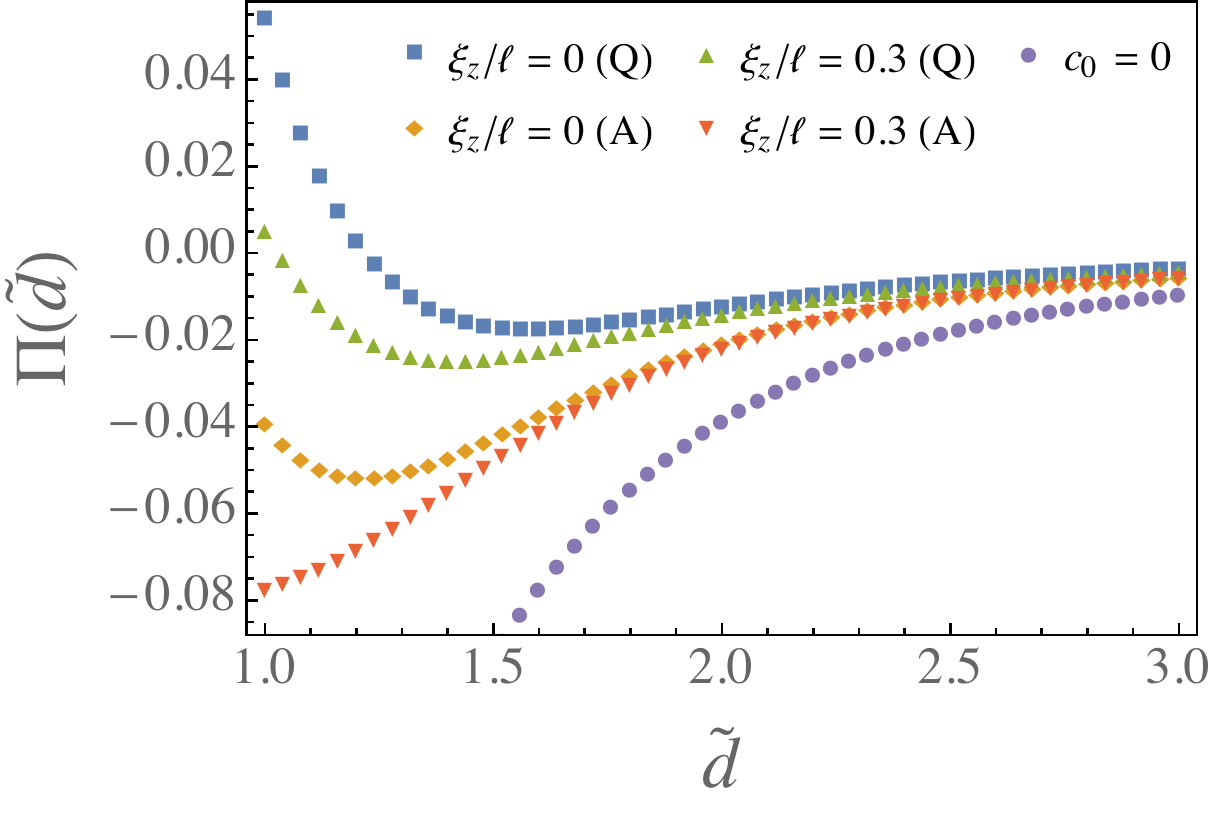}
 		\caption{Rescaled interaction pressure ${\Pi}(\tilde d)$ as a function of the rescaled separation $\tilde d=d/\lambda$ in the two cases of  annealed (A) and quenched (Q) disorder with fixed $\lambda/a_0=2$,  $\lambda/\ell =10$ and $\beta K c_0/\ell= 0.5$, and with and without spatial in-plane correlations, $\xi_z/\ell=0.3$ and 0, respectively. The disorder-free case is shown by purple circles (bottom-most data set). 
					}
 		\label{fig:fig4}
 	\end{center}
\end{figure}

 In Fig.~\ref{fig:fig4}, the interaction pressure profiles of the annealed and quenched cases are compared for spatially uncorrelated ($\xi_z/\lambda=0$) and correlated ($\xi_z/\lambda=0.3$) disorder models  with fixed rescaled variance of $\beta K c_0/\ell=0.5$. As seen, quenched disorder is more efficient in reducing the absolute strength of the fluctuation-induced force across all inter-substrate separations. In addition, the presence of in-plane substrate disorder correlations invariably tends to suppress the disorder effects (compare with the disorder-free case).  As an interesting counter point, the typical difference between the quenched and annealed cases  in a nematic cell  \cite{faar} is  smaller than those obtained here. Also, as opposed to the nematic case \cite{faar}, the annealed disorder   in the sm-A system leads to a more strongly attractive interaction profile than the quenched one across all inter-substrate separations.

\section{Conclusion}
\label{sec:con}

The beating rhythm of the vacuum at absolute zero---the quantum electromagnetic field fluctuations---gives rise to a fluctuation-induced Casimir force between two ideally polarizable plates \cite{Casimir}. At finite temperatures a thermal component upgrades the Casimir interaction to a van der Waals interaction \cite{Parsegian}.  In general, the thermal component to fluctuation induced interactions in confined geometries operates also for critical fluctuating fields that are not electromagnetic in nature, as already hypothesized in the classical work of Ref. \cite{Dzyaloshinski}. These fields can describe various physical systems either close to a critical point or with infinite range potentials implying infinite range correlation functions \cite{Kardar}.

Among the systems with effective infinite range potentials, elastic fluctuations in liquid crystalline systems have been a particularly fertile field of research. Fluctuations of the layer spacing in a confined sm-A liquid crystal film were shown to imply the presence of a long-range thermal pseudo-Casimir interaction in the case of a bookshelf geometry configuration \cite{FAR}. In the limit of small displacements from the thermodynamic equilibrium, the fluctuations can be described by a local displacement field $u({\mathbf r}_{\!_\perp}, y)$, at each transverse ${\mathbf r}_{\!_\perp}=(x,z)$ as well as perpendicular spatial coordinate $y$. For each ${\mathbf r}_{\!_\perp}$ an infinite number of Fourier modes with wave numbers ${\mathbf p}_{\!_\perp}=(p_x,p_z)$ contribute to the fluctuation spectrum. As is usual for Fourier decompositions, 
the Hamiltonian of the system can be diagonalized in the ${\mathbf p}_{\!_\perp}$-space. While $p_x$ changes in the interval $(-\infty, +\infty)$, changes in $p_z$ remain confined to the first Brillouin zone,  $p_z\in [-\pi/a_0,\pi/a_0)$, with $a_0$ being the layer periodicity. The Fourier decomposed Hamiltonian then contains second order derivatives in the perpendicular spatial coordinate, making it interesting also from a more formal point of view \cite{Dean-FI}. The fluctuation free energy can be evaluated straightforwardly and studied as a function of the size of the film, $d$, in the $y$ direction, showing interesting variation in sign as well as magnitude as a function of model parameters and geometry. 

In summary, the fluctuations and finite-size effects together lead to a force which is the derivative of the corresponding confined thermal fluctuation free energy with respect to the thickness of the film, assuming that the density of the material in the film remains constant. Presence of disorder in the easy axis of the anchoring, that can be formally analyzed within the standard replica {\em Ansatz} approach,  wroughts important modifications on the pseudo-Casimir force. Quenched disordered anchoring field, leading to the coupling between different replicas of the deformation field $u_a$, with $a=1,\cdots, k$ the index of the replica, gives rise to an extra additive contribution to the pseudo-Casimir force in a disorder-free configuration, suppressing the strength of the net interaction force when compared with the disorder-free case. The same trend   is observed also in the presence of annealed disorder. In this latter case, the disorder effectively weakens the anchoring strength and thus reduces the pseudo-Casimir force through a modification of the structural parameters.  It is finally interesting to note that the forces induced by sm-A fluctuations are found to be significantly larger than those produced in the case of nematic films \cite{faar}.  This appears to be a simple consequence of the fact that in the former case the unit of length is the penetration length $\lambda$, comparable to  the molecular layer length $a_0$, whereas in nematics the unit of length is usually the surface extrapolation length $\ell$, which varies from molecular length $a_0$ and then all the way up to $10^{3}a_0$. This indeed makes the pseudo-Casimir interactions mediated by sm-A films more interesting and experimentally easier to detect.



\begin{thebibliography}{99}

\bibitem{Woods} L. M. Woods, D. A. R. Dalvit, A. Tkatchenko, P. Rodriguez-Lopez, A. W. Rodriguez, and R. Podgornik, Rev. Mod. Phys. {\bf 88}, 045003 (2016).
\bibitem{Parsegian} V. A. Parsegian, {\em Van der Waals Forces: A Handbook for Biologists, Chemists, Engineers, and Physicists} (Cambridge University Press, Cambridge, 2006).
\bibitem{French} R.H. French, V. A. Parsegian, R. Podgornik, et al., Rev. Mod. Phys. 2010, {\bf 82}, 1887.

\bibitem{Bordag} Bordag, M., G. L. Klimchitskaya, U. Mohideen, and V. M. Mostepanenko, {\em Advances in the Casimir effect} (Oxford University Press, Oxford, UK, 2009).
\bibitem{Casimir} H. B. G. Casimir, Proc. Kon. Ned. Akad. Wet. {\bf 51}, 793 (1948).
\bibitem{Rahi} S. J. Rahi, T. Emig, N. Graham, R. L. Jaffe, and M. Kardar, Phys. Rev. D {\bf 80}, 085021 (2009).
\bibitem{Somers} D. A. T. Somers, J. L. Garrett, K. J.  Palm, and J. N. Munday, Nature {\bf 564}, 386 (2018).
\bibitem{zumer} S. {\v Z}umer, Nature {\bf 564}, 350 (2018).
\bibitem{Fisher} M.E. Fisher and P.-G. de Gennes, C. R. Acad. Sci. Ser. B {\bf 287}, 207 (1978).
\bibitem{Dietrich} C. Hertlein, L. Helden, A. Gambassi, S. Dietrich, and C. Bechinger, Nature {\bf 451}, 172 (2008). 
\bibitem{Ajdari} A. Ajdari, B. Duplantier, D. Hone, L. Peliti, and J. Prost, J. Phys. II {\bf 2}, 487 (1992). 
\bibitem{Kardar} M. Kardar and R. Golestanian, Rev. Mod. Phys. {\bf 71}, 1233 (1999).

\bibitem{de-gennes} P. G. de Gennes and J. Prost, {\em The Physics of Liquid Crystals} (Oxford Science Publications, Oxford, 1995).

\bibitem{sndhp} J. Sarabadani, A. Naji, D. S. Dean, R. R. Horgan, and R. Podgornik, J. Chem. Phys. {\bf 133}, 174702 (2010).
\bibitem{ddnp} D. S. Dean, J. Dobnikar, A. Naji, and R. Podgornik, {\em Electrostatics of Soft and Disordered Matter} (Singapore, Pan Stanford Publishing, 2014). 
\bibitem{radzihovsky} T. Bellini, L. Radzihovsky, J. Toner, and N. A. Clark, Science {\bf 294}, 1074 (2001).
\bibitem{faar} F. Karimi Pour Haddadan, A. Naji, A. Khame Seifi, and R. Podgornik, J. Phys: Condens. Matter {\bf 26}, 075103 (2014); {\em ibid} {\bf 26}, 179501 (2014).
\bibitem{fanr} F. Karimi Pour Haddadan, A. Naji, N. Shirzadiani, and R. Podgornik, J. Phys: Condens. Matter {\bf 26}, 505101 (2014).

\bibitem{Dean-FI} D. S. Dean, B. Miao, R. Podgornik, submitted (2019).

\bibitem{FAR} F. Karimi Pour Haddadan, A. Naji, and R. Podgornik, Soft Matter {\bf 15}, 2216 (2019).
	\bibitem{RP} A. Rapini and M. Papoular, J. Phys. Coll. {\bf 30}, C4-54 (1969).
	\bibitem{radzihovsky1} Q. Zhang and L. Radzihovsky, Phys. Rev. E {\bf 81}, 051701 (2010).
	\bibitem{radzihovsky2} L. Radzihovsky and Q. Zhang, Phys. Rev. Lett. {\bf 103}, 167802 (2009).
\bibitem{lubensky} P. M. Chaikin and  T. C. Lubensky, {\em Principles of Condensed Matter Physics} (Cambridge University Press, Cambridge, 1995).

\bibitem{dotsenko1}
V. Dotsenko, {\em Introduction to the Replica Theory of
Disordered Statistical Systems} (Cambridge University Press, New York, 2001).

\bibitem{dotsenko2}
 V. Dotsenko, {\em An Introduction to the Theory of Spin Glasses and
Neural Networks} (World Scientific,  Singapore, 1994).

\bibitem{partial}
Y. Sh. Mamasakhlisov, A. Naji, and R. Podgornik, J. Stat. Phys. {\bf 133}, 659 (2008).

\bibitem{kleinert} H. Kleinert, J. Math. Phys. {\bf 27}, 3003 (1986).
\bibitem{handbook} C.  Grosche and F. Steiner, {\em Handbook of Feynman Path Integrals} (Springer-Verlag, Berlin, 1998). 

\bibitem{LC3a}
H. Li and M. Kardar, Phys. Rev. Lett. {\bf 67}, 3275 (1991). 
\bibitem{LC3b}
H. Li and M. Kardar, Phys. Rev. A {\bf 46}, 6490 (1992).
\bibitem{radzihovsky3}Q. Zhang and L. Radzihovsky, EPL {\bf 98}, 56007 (2012).
\bibitem{Dzyaloshinski}  I.~E. Dzyaloshinskii, E.~M  Lifshitz, and L.~P. Pitaevskii, Adv. Phys. {\bf 10}, 165 (1961).	

\end{thebibliography}
\end{document}